\documentclass[10pt]{article}
\usepackage{a4,epsfig,amsmath,amssymb,cite}

\begin{document}

\begin{center}
   \centering{\large\bf A protein model exhibiting three folding transitions}\\

  \vspace*{1cm}
  \centering{Audun Bakk\\
 {\it Department of Physics, Norwegian University of Science and 
     Technology, NTNU, N-7491 Trondheim, Norway}\\
  \vspace*{0.5cm}
  Alex Hansen\footnote{Permanent Address: Department of Physics, Norwegian University of Science and Technology, NTNU, N-7491 Trondheim, Norway} and Kim Sneppen\\
  {\it NORDITA and Niels Bohr Institute, Blegdamsvej 17, DK-2100 Copenhagen, Denmark}}\\
  
  \vspace*{0.5cm}
  \centering{(\today)}
\end{center}
\vspace*{0.5cm}

\begin{abstract}
We explain the physical basis of a model for small globular proteins with water interactions. The water is supposed to access the protein interior in an ``all-or-none'' manner during the unfolding of the protein chain. As a consequence of this mechanism (somewhat speculative), the model exhibits fundamental aspects of protein thermodynamics, as cold, and warm unfolding of the polypeptide chain, and hence decreasing the temperature below the cold unfolding the protein folds again, accordingly the heat capacity has three characteristic peaks. The cold and warm unfolding has a sharpness close to a two-state system, while the cold folding is a transition where the intermediate states in the folding is energetical close to the folded/unfolded states, yielding a less sharp transition. The entropy of the protein chain causes both the cold folding and the warm unfolding.
\end{abstract}


\section{Introduction}
\label{sec:1}

In order to have a precise function in the biological ``machinery'', it is important for proteins to have an unique conformation at physiological temperatures. This is termed the {\it native} state. Anfinsen\,\cite{Anfinsen:73} proved in his famous experiment with ribonuclease the important fact that the folding of the polypeptide chain is thermodynamically determined.

One simplified view of protein folding is that the protein is supposed to follow a specific {\it folding pathway} of conformations in a descending landscape of Gibbs free energy\,\cite{Baldwin:99a,Baldwin:99b,Wang:95,Hansen:98a,Hansen:99,Hansen:98b,Bakk:00a,Bakk:00b}. This is a picture of a folding protein that is forced to follow a specific ``path'' of successive conformational steps of increasing structural order. We will use this pathway assumption in this paper.

A protein in physiological environments (pH, ionic strength {\it etc.}), and temperatures is packed in a very compact way. It is then termed {\it folded}. An increase of the temperature will eventually denaturate the protein, {\it i.e.}\hspace{0.13cm}it {\it unfolds}. Other ways to unfold the protein are for instance to change the pressure, denaturant concentration or the pH. The fact that proteins also unfold at low temperatures, termed as {\it cold unfolding}\,\cite{Privalov:90,Chen:89}, makes the system very unusual. A major difficulty in experiments of cold unfolding is that the temperature is around and below the freezing point of water. In a frozen aqueous solution, one cannot observe any conformational transitions\,\cite{Creighton:92}.

A general feature of small globular proteins is that they {\it thermodynamically} seems to unfold in an ``all-or-none'' manner. This means that they unfold cooperatively without noticeable intermediates\,\cite{Privalov:74,Privalov:79,Makhatadze:93,Jackson:91,Bae:95,Griko:94}, with a deviation from a two-state system not exceeding 5\,\%. The deviation from a single macroscopic system can be explained by presence of unstable intermediates\,\cite{Bakk:00a,Privalov:79,Dommersnes:00}. It is worth noting that all these experiments have been done only for the warm unfolding. The occurrence of intermediate states in larger proteins\,\cite{Baldwin:99a,Baldwin:99b,Shortle:96} is not a contradiction to the two-state behavior in the experiments in Refs.\,\cite{Privalov:74,Privalov:79,Makhatadze:93,Jackson:91,Bae:95,Griko:94}, because the latter only considers small globular proteins. 

\begin{figure}
   \caption{}
   \vspace{2cm}
   \centering{\epsfig{figure=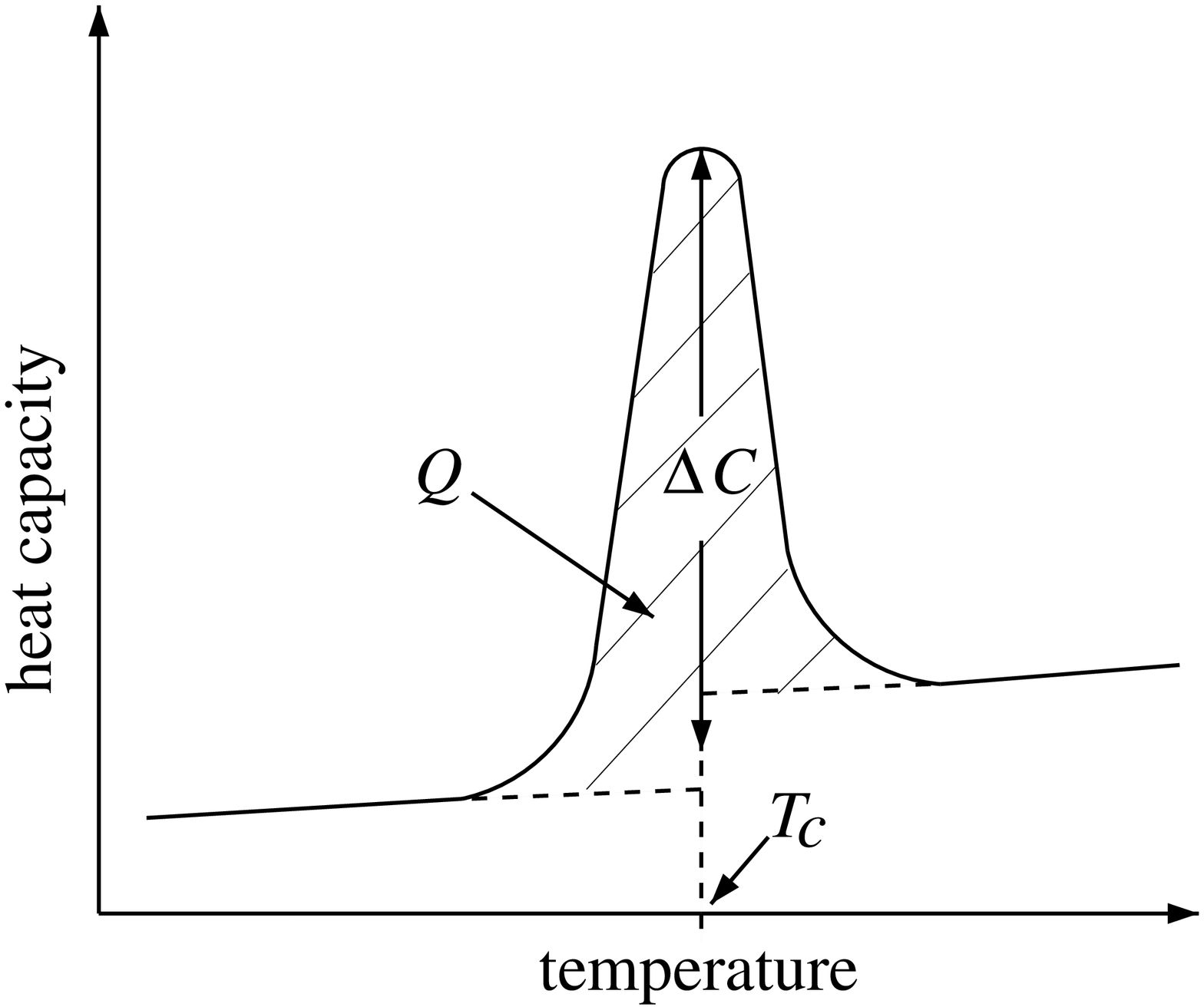,width=\linewidth}}
   \label{fig:1}
\end{figure}

The van't Hoff enthalpy relation (for heat of reaction)\,\cite{Privalov:79,Callen:85}  
\begin{equation}
  \label{vH}
  \Delta H=\alpha\,k_B\,T_c^2\,\frac{\Delta C}{Q}\quad ,
\end{equation}
is a powerful way to quantify the {\it sharpness} of a smoothed out first order phase transition. As shown in Fig.\,\ref{fig:1}, $T_c$ is the transition temperature (at the middle of the peak), $Q$, which is the same as $\Delta H$ (no pressure), is the released energy (latent heat), and $\Delta C$ is the peak height of the transition. $\alpha$ is a dimensionless proportionality factor. For a given $\Delta H$ and $Q$, then the value of $\alpha$ is inversely proportional to $\Delta C$. In this respect a smaller $\alpha$ corresponds to a sharper transition. 

\vspace{0.5cm}
In this article we will explain the physical basis of a protein model, that reformulates the water interactions proposed in earlier models by Hansen {\it et al.}\,\cite{Hansen:98a,Hansen:99} and Bakk {\it et al.}\,\cite{Bakk:00a,Bakk:00b}. We will compare thermodynamical quantities, as the heat capacity, to experiments. The protein is also investigated in a temperature region below accessible experimental data.
\section{Modeling the protein}
\label{sec:2}

\subsection{The polypeptide chain}
\label{sub:21}
The polypeptide chain is modeled as in earlier articles by Hansen {\it et al.}\,\cite{Hansen:98a,Hansen:99,Hansen:98b} and Bakk {\it et al.}\,\cite{Bakk:00a,Bakk:00b}, where the protein is supposed to follow a pathway as described in Section\,\ref{sec:1}. The protein is equipped with $N$ contact points, which we here call {\it nodes}. A node in contact means that the protein has a ``correct'' conformation on the folding pathway, and we say that the node is folded. Due to the fact that a protein is a complex 3-dimensional system, a folded node likely has non-local contacts with respect to the amino acid sequence in the polypeptide chain.

Each node is assigned only two energies $-\epsilon_0$ or 0 \cite{Bryngelson:87,Bryngelson:90}. By using the binary variables $\phi_i\in\{0,1\}$, the energy associated to each individual node is written
\newline  $E_i=-\epsilon_0\,\phi_1\phi_2\cdots\phi_{i-1}\phi_i\,.$ The value $\phi_i=0$ means an unfolded node, while $\phi_i=1$ is equivalent to a folded node. The product term meets the assumption about a folding pathway, because $E_i=-\epsilon_0$ only if all nodes before $i$ is folded, in addition to a folded node $i$ itself. For a system of $N$ nodes the Hamiltonian of the polypeptide chain (vacuum energy of the protein) is
\begin{equation}
   \label{H_c}
   {\cal H}_c=-\epsilon_0(\phi_1 + \phi_1\phi_2  
           + \cdots+\phi_1\phi_2\cdots\phi_N)\quad .
\end{equation}  
Folding of node $i$ is only proper in the case of an unique conformation before this folding step. According to Eq.\,\ref{H_c}, every attempt to fold $i$ while $j$ (describing one or several nodes $<i$) is unfolded will not gain energy in the system, because then the protein is supposed to be in an energetical unfavorable conformation.

The unfolded protein has more degrees of freedom relative to the folded, because the unfolded polypeptide backbone will have rotational freedom. We incorporate this by assigning each unfolded node $f$ degrees of freedom. The parameter $f$ is interpreted as the relative increase in the degrees of freedom for an unfolded node compared to a folded node.   

\subsection{Water interactions}
Interactions between water and protein surface is important. Proteins are during the evolution ``designed'' to interact with water, simply because they are exposed to water {\it in vivo}. Makhatadze and Privalov\,\cite{Makhatadze:95} states that in sum hydration effects destabilize the native state, and decreasing temperature implies increasing destabilizing action. The water that access the hydrophobic protein interior during unfolding is supposed to obtain an ``ice-like'' structure around the apolar surfaces. Hence, this structured water will both decrease the entropy and the energy compared to ``free'' water\,\cite{Privalov:92}, and thus impacts the thermodynamics of the system.

Hansen {\it et al.}\,\cite{Hansen:98a,Hansen:99} proposed a simple model where each water molecule interacts with a node, and not with other water molecules, by a ``ladder'' of $g$ equidistant energies accessible  
\begin{equation}
  \label{ladder}
  \omega_{i}=
    \begin{cases}
      -\varepsilon_w +(g-1)\delta\\
      \hspace{0.828cm}\vdots\\
      -\varepsilon_w +2\delta\\
      -\varepsilon_w +\delta\\
      -\varepsilon_w\quad ,
    \end{cases}
\end{equation}
which we will also apply in the model considered in this text. The interpretation of  $\omega_{i}$ is the energy difference between a ``structured'' water molecule, associated to the unfolded parts of the protein, and a ``free'' water molecule in the bulk.

The observable states in a small globular protein is either the unfolded ($\phi_1=0$), with water bounded to the surface that uncovers during unfolding of the protein, or the folded state ($\phi_1\cdots\phi_N =1$) with no water in the protein interior. No intermediate states are detected for small globular proteins\,\cite{Privalov:96}, hence one cannot know for sure how the water enters the protein interior during the unfolding. Hansen {\it et al.}\,\cite{Hansen:98a,Hansen:99} and Bakk {\it et al.}\,\cite{Bakk:00a,Bakk:00b} have earlier only considered that the amount of water interactions increase proportional to the number of unfolded nodes, and with that the contact energy of the chain. In this paper we study, as a more speculative assumption, the case when a macroscopic contribution of water enters the protein surface when the last node is unfolded.

We note that Eq.\,\ref{ladder} is the quantized energy levels of a magnetic dipole in an external field. In the continuum limit where $g\rightarrow\infty$ (with $g\,\delta$ finite), a classical magnetic dipole in an external field is obtained, and this again is analogous to an electrical dipole in an external electrical field. The dipolar water molecules are exposed to an electrical field from the permanent, and induced charges on the protein surface, thus Eq.\,\ref{ladder} is a representation of that.   
 
By using the same notation as in Eq.\,\ref{H_c}, we propose the Hamiltonian that corresponds to the water-protein interactions
\begin{equation}
  {\cal H}_w=(1-\phi_1\phi_2\cdots\phi_N)\,
             (\omega_1+\omega_2+\cdots+\omega_M)\quad ,
\end{equation}
where $M$ is the number of water molecules.
The folded protein is a highly ordered and dense packed structure where no water can access the interior. Due to Eq.\,\ref{H_c}, unfolding of the last node $(\phi_N=0)$ implies a less dense pacing of the protein, and the cavities are now supposed to be big enough to let water access the interior of the protein. The next step, unfolding of node $N-1$, implies likely an even lesser dense packing, and allows more water in the protein interior. We assume in this text that the water entering upon unfolding of node $N-1$, will not interact with the protein surface, because it is regarded as a second layer of ``free''water. Cohn and Edsall\,\cite{Cohn:43} states that roughly a monolayer of water is bounded to the protein, implying that the protein is only interacting with the first monolayer, thus the second, and third {\it etc.}\hspace{0.13cm}water layers, successively entering the protein during unfolding, are regarded as ``free water''. Hence, according to the latter possible (but somewhat speculative) explanation of how the water access the apolar interior of the protein,  unfolding of nodes $i<N$ does not contribute energetical to the water Hamiltonian (${\cal H}_w$) and thus not to the thermodynamics.  

\subsection{The statistical framework}
The system Hamiltonian (${\cal H}$) describing both chain specific energy (${\cal H}_c$) and water interactions (${\cal H}_w$) is
\begin{equation}
 \begin{split}
   {\cal H}={\cal H}_c+{\cal H}_w=&-\epsilon_0\,
            (\phi_1+\phi_1\phi_2 \cdots+\phi_1\phi_2\cdots\phi_N)\\
            &+(1-\phi_1\phi_2\cdots\phi_N)(\omega_1+\omega_2+\cdots +\omega_M)
 \end{split}
\end{equation}
Let now $Z_i$ be term number $i$ in the partition function which corresponds to folding of all nodes $\leq i$ (pathway assumption), thus
\begin{equation}
   \label{Z_i1}
   Z_i=f^{N-i}\,e^{i\epsilon_0\beta}\,
       \left( e^{\varepsilon_w\beta}\frac{1-e^{-g\delta\beta}}
                                     {1-e^{-\delta\beta}}\right)^M
        \quad \; (i<N)\quad .
\end{equation}
$\beta\equiv 1/T$ is a rescaled inverse absolute temperature in which the Boltzmann constant is absorbed. 
$Z_0$ is the term where all nodes are zero, {\it i.e.}\hspace{0.13cm}a complete unfolded protein, while $Z_N$ corresponds to a folded protein. The factor $f^{N-i}$ in Eq.\,\ref{Z_i1} is the degrees of freedom in the polypeptide chain that is available in the $N-i$ unfolded nodes. $e^{i\epsilon_0\beta}$ is the Boltzmann factor from $i$ contact energies $-\epsilon_0$ in the polypeptide chain. The last term in brackets is simply the sum over all distinct levels in one water molecule raised to the power of the number of water molecules $M$ bounded to the unfolded parts of the protein.   
We assume that $\delta\beta\ll 1$ ({\it i.e.}\hspace{0.13cm}$g\rightarrow\infty$), which is equal to an infinite small level spacing in Eq.\,\ref{ladder}. A first order Taylor expansion of the denominator in Eq.\,\ref{Z_i1} yields
\begin{equation}
   \label{Z_i2}
   Z_i=f^N\,\left( \frac{e^{\varepsilon_w\beta}}{\delta\beta}\right)^M
       \,e^{i(\epsilon_0\beta-\ln{f})}\quad \; (i<N)\quad ,
\end{equation}
assuming $1-e^{-g\delta\beta}\approx 1$ when $g\rightarrow\infty$.
The last term in the partition function ($Z_N$) corresponds to a complete folded protein, where there are  $g^M$ degrees of freedom from $M$ unbounded water molecules and $N$ contact energies $-\epsilon_0$, hence
\begin{equation}
   \label{Z_N}  
   Z_N=g^M\,e^{N\epsilon_0\beta}\quad .
\end{equation}

By summing up the $Z_i$ terms in Eqs.\,\ref{Z_i2} and \ref{Z_N}, we obtain the partition function
\begin{equation}
   \label{Z}
   Z=\sum_{i=0}^{N}Z_i
    =f^N\,g^M\left[ \left( \frac{a\,e^{\mu\beta}}{\beta}\right)^M\,
                    \frac{1-r^N}{1-r}
                    +r^N\right]\quad ,
\end{equation}
where $r\equiv e^{\beta-\ln{f}}$, $a\equiv\epsilon_0/(g\delta)$ and $\mu\equiv \varepsilon_w/\epsilon_0$. The inverse temperature is here rescaled by $\epsilon_0\,\beta\rightarrow\beta$. The parameter $\mu$ measures the strength of the water interactions relative the chain contact energy, thus $\mu$ is interpreted as an {\it effective chemical potential}. Changing $\mu$ means adding denaturants, changing pH or salt concentration {\it etc.} 

The order parameter $n$ in this system measures the degree of folding, {\it i.e.}\hspace{0.13cm}the mean number of folded nodes divided by $N$
\begin{equation}
  \label{n}
  n=\frac{\sum_{i=0}^{N}i\, Z_i}{Z}
   =\frac{1}{N}\,
   \frac{\left( \frac{a\,e^{\mu\beta}}{\beta}\right)^M\,
          \frac{(N-1)\,r^{N+1}-N\,r^N+r}{(1-r)^2}+N\,r^N}
        {\left( \frac{a\,e^{\mu\beta}}{\beta}\right)^M\,
          \frac{1-r^N}{1-r}+r^N}\quad .
\end{equation}
$n=0$ corresponds to an unfolded protein, while $n=1$ is interpreted as a completely folded protein.
\section{Calculations and discussion}
The heat capacity is 
$C=\beta^2\cdot \partial^2 (\ln{Z})/\partial\beta^2$. Fig.\,\ref{fig:2} shows a typical plot of the heat capacity $C(T)$ with three peaks (numbered 1, 2 and 3 from left). These characteristic peaks corresponds to three critical transition temperatures: $T_1$, $T_2$ and $T_3$, measuring the temperatures at the respective peak maxima. The corresponding order parameter $n(T)$ in Fig.\,\ref{fig:3}, calculated from Eq.\,\ref{n}, shows that the protein is essential folded for $T<T_1$ and $T_2<T<T_3$, while the protein is nearly unfolded in the temperature intervals $T_1<T<T_2$ and $T>T_3$. From this picture it is reasonable to state that the physiological temperature interval is between peak 2 and 3. Accordingly, with reference to this temperature region, we call peak 1 for {\it cold folding} and peak 2 and 3 respectively for {\it cold} and {\it warm unfolding}. Peak 2 and 3 are both observed in experiments\,\cite{Privalov:90,Chen:89} and are also seen in the model of Hansen {\it et al.}\,\cite{Hansen:98a,Hansen:99} and Bakk {\it et al.}\,\cite{Bakk:00a,Bakk:00b}. The model considered in this paper has, in addition to the cold and warm unfolding, the peculiarity of {\it cold folding}. 

Experiments on cold unfolding are very difficult because most proteins unfolds below the freezing point of water. Chen and Chellman\,\cite{Chen:89} and Privalov {\it et al.}\,\cite{Privalov:86} have all done experiments where the cold unfolding temperature is elevated by denaturants, but denaturants make the interpretation of the data very difficult. However, Privalov\,\cite{Privalov:90} did experiments in super cooled water, which is easier to interpret. Unfortunately he was not able to detect the sharpness of the cold unfolding, and not at all the heat capacity below the cold unfolding. This means that our model {\it may} predict a cold folding transition at a temperature below the cold unfolding transition.

For temperatures below the cold folding ($T\rightarrow 0$) analysis of the model gives $n\approx 0.99$, {\it i.e.}\hspace{0.13cm}only the last node is unfolded, and corresponds to the global energy minimum. From Fig.\,\ref{fig:2} it is seen that $T_1\approx 1.45$. In Eq.\,\ref{Z} the critical \newline $r\equiv e^{1/T\,-\,\ln{f}}=1$, caused by the contact energies of the polypeptide chain, implies $T_1=1/\ln{2}\approx 1.44$. Hence, this is nothing but a transition initiated of the chain entropy. An increase of the temperature from $T_1$ takes the protein through a nearly unfolded state, whereupon the protein folds again at $T_2\approx 1.8$. The temperature is now so high that the energy of water (Eq.\,\ref{ladder}) is small  (thermal exited) compared to the chain contact energy $\epsilon_0$, thus the protein prefers to fold again. Further increase of the temperature causes warm unfolding at $T_3\approx 3.0$, because then the entropy of the chain again dominates the Gibbs free energy. It is interesting to note that the entropy of the chain causes two transitions, the cold folding and warm unfolding.        

We now turn our interest to the sharpness of the transitions, {\it i.e.}\hspace{0.13cm}the value of the parameter $\alpha$ in the van't Hoff enthalpy relation (Eq.\,\ref{vH}). For $M=200$ is $\alpha\approx 4$ both for the cold and warm unfolding. This means that the protein is thermodynamically regarded as a two-state system that folds in an ``all-or-none'' manner. Privalov\,\cite{Privalov:79} has measured  $\alpha =4.0$ for the warm unfolding. As far as we know there are no experiments on the sharpness of the cold unfolding, but Privalov\,\cite{Privalov:90} indicates a sharpness for the cold unfolding as well, thus according to our model. The cold folding transition has $\alpha\approx 12$. This value is typical for a transition where one has small energy differences between the folded/unfolded states and the intermediate states. Remember that the ``folded'' state for $T<T_1$ is actually the first node unfolded, thus the unfolding will essential depend on the polypeptide chain with the Hamiltonian in Eq.\,\ref{H_c}, which can be shown corresponds to $\alpha =12$\;\cite{Bakk:00a,Dommersnes:00}. 

Finally we note the consequence of a decreased $\mu$ is an increasing separation between the cold and warm unfolding as seen in Fig.\,\ref{fig:4}. This makes sense because a smaller $\mu$ is equivalent to a relatively smaller $\varepsilon_w$ compared to $\epsilon_0$ (see Eq.\,\ref{Z}), {\it i.e.}\hspace{0.13cm}it is less favorable for the protein to be bounded to water. The consequence is that the protein prefers to be folded in a larger temperature interval, in where the water is expelled to the bulk. However, the transition temperature $T_1$ is not changed because this transition is given by the value $T_1=1/\ln{f}$. It is also seen that a smaller $\mu$ is qualitatively equivalent to a smaller $a$. 

An increase in $M$ is the same as a decrease in $N$, because then the water becomes more important relative to the chain, and will again allow a broader separation between $T_2$ and $T_3$. The broader separation is also seen for a decreasing $f$, because this is equivalent to a larger $M$. 

Further increase of $\mu$ will eventually merge peak 1 and 2. It is interesting to note that $\alpha=4$ for the merged peaks, because then the transition is energetically dominated of the $M$ water molecules which caused the transition at $T_2$ in Fig.\,\ref{fig:2}.

In sum the qualitative change from Fig.\,\ref{fig:3} to Fig.\,\ref{fig:4}, by a decreasing $\mu$, is also obtained by an decrease of $a$, $f$ or $N$ or an increase of $M$.

\section{Conclusion}
\label{sec:4}

We have in this paper studied a protein model with water interactions. The model is based on earlier models by Hansen {\it et al.}\,\cite{Hansen:98a,Hansen:99} and Bakk {\it et al.}\,\cite{Bakk:00a,Bakk:00b}. In contrast to these similar models, where the water amount was supposed to increase linearly to the degree of unfolding of the polypeptide chain, we have, with a more speculative assumption, studied the situation where a macroscopic amount of water access the protein interior during unfolding of the last node is the only contribution to the water-protein Hamiltonian.

With reference to physiological temperatures we find that the protein exhibits cold and warm unfolding transitions, which is an experimental fact\,\cite{Privalov:90,Chen:89}. These transitions are associated by a sharpness indicating, from at thermodynamical point of view, a two-state system, which is also experimental established\,\cite{Privalov:74,Privalov:79,Makhatadze:93,Jackson:91,Bae:95,Griko:94}. Decreasing the temperature further below the cold unfolding region the protein folds again (cold folding). This folding, caused by the chain entropy, has a less sharp transition, which corresponds to a transition where the intermediate folding energies does not differ significant from the folded/unfolded energies. In sum the model exhibits three unfolding/folding transitions.

It is interesting to note that {\it both} the cold folding and the warm unfolding is due to the polypeptide chain entropy.    

\newpage
\vspace{2cm}
\noindent
{\large\bf Acknowledgements}

\vspace{0.5cm}
\noindent
A. B. thanks the Norwegian Research Council for financial support. A. H. thanks NORDITA and Niels Bohr Institute for warm hospitality and support. We thank J. S.  H\o ye for enthusiastic and enlightening discussions. 


\newpage
\vspace{2cm}
\noindent
{\large\bf Figure captions}

\vspace{0.5cm}
\noindent
{\bf Fig. 1.} Schematic illustration of the heat capacity around an unfolding transition showing the parameters in the van't Hoff enthalpy relation (Eq.\,\ref{vH}). $T_c$ is the transition temperature, $Q$ (area of the peak) is the released energy (latent heat) and $\Delta C$ is the peak height of the transition.

\vspace{0.5cm}
\noindent
{\bf Fig. 2.} Heat capacity $C(T)$ with the parameters $a=0.077$, $\mu=3.3$, $f=2$, and $N\,$=$\,M\,$=$\,200$, showing three peaks. With reference to the temperature region between peak 2 and 3 (physiological temperatures) we call the transitions: 1) {\it cold folding}, 2) {\it cold unfolding} and 3) {\it warm unfolding}.

\begin{figure}
   \caption{}
   \vspace{2cm}
   \centering{\epsfig{figure=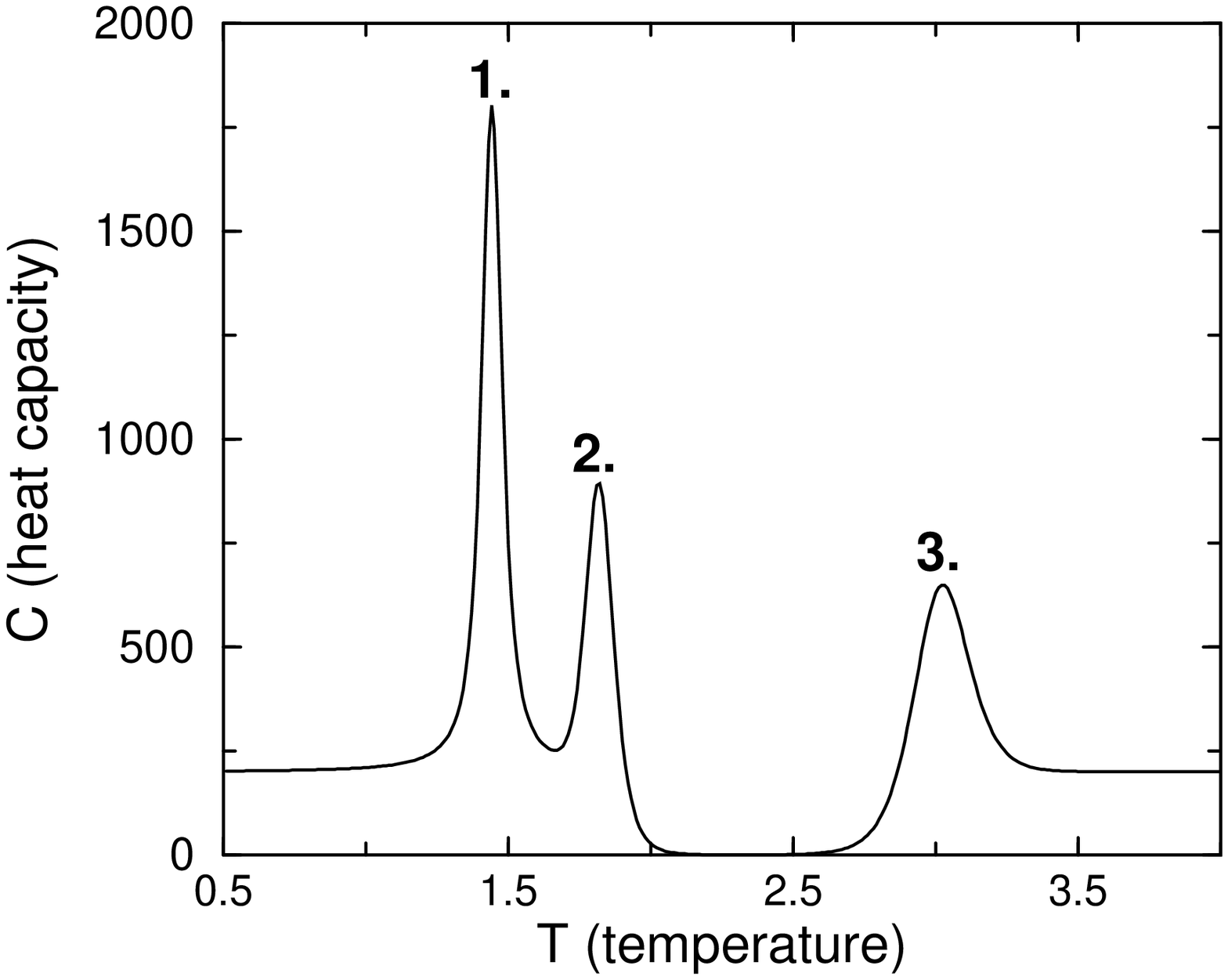,width=\linewidth}}
   \label{fig:2}
\end{figure}

\begin{figure}
   \caption{}
   \vspace{2cm}
   \centering{\epsfig{figure=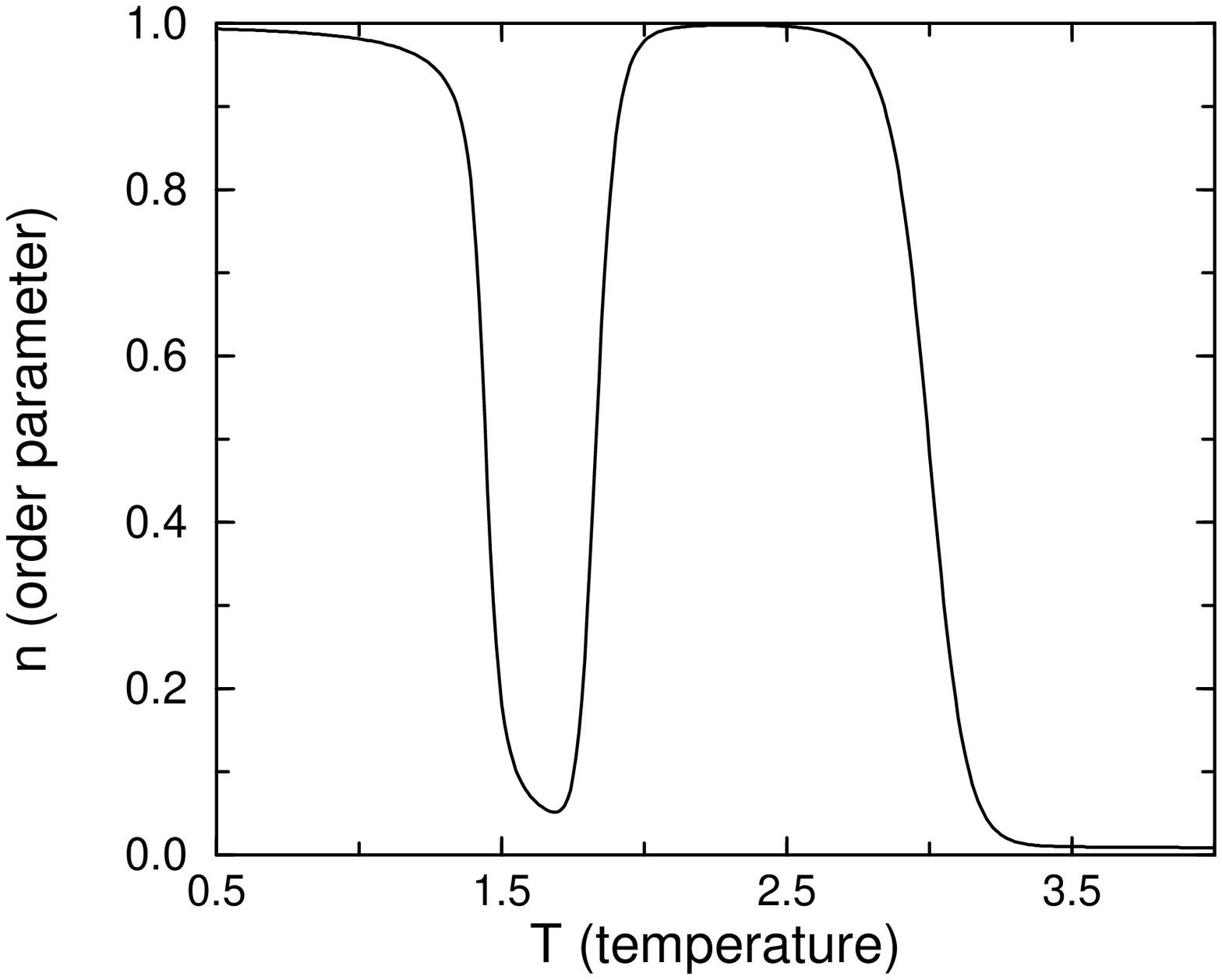,width=\linewidth}}
   \label{fig:3}
\end{figure}

\begin{figure}
   \caption{}
   \vspace{2cm}
   \centering{\epsfig{figure=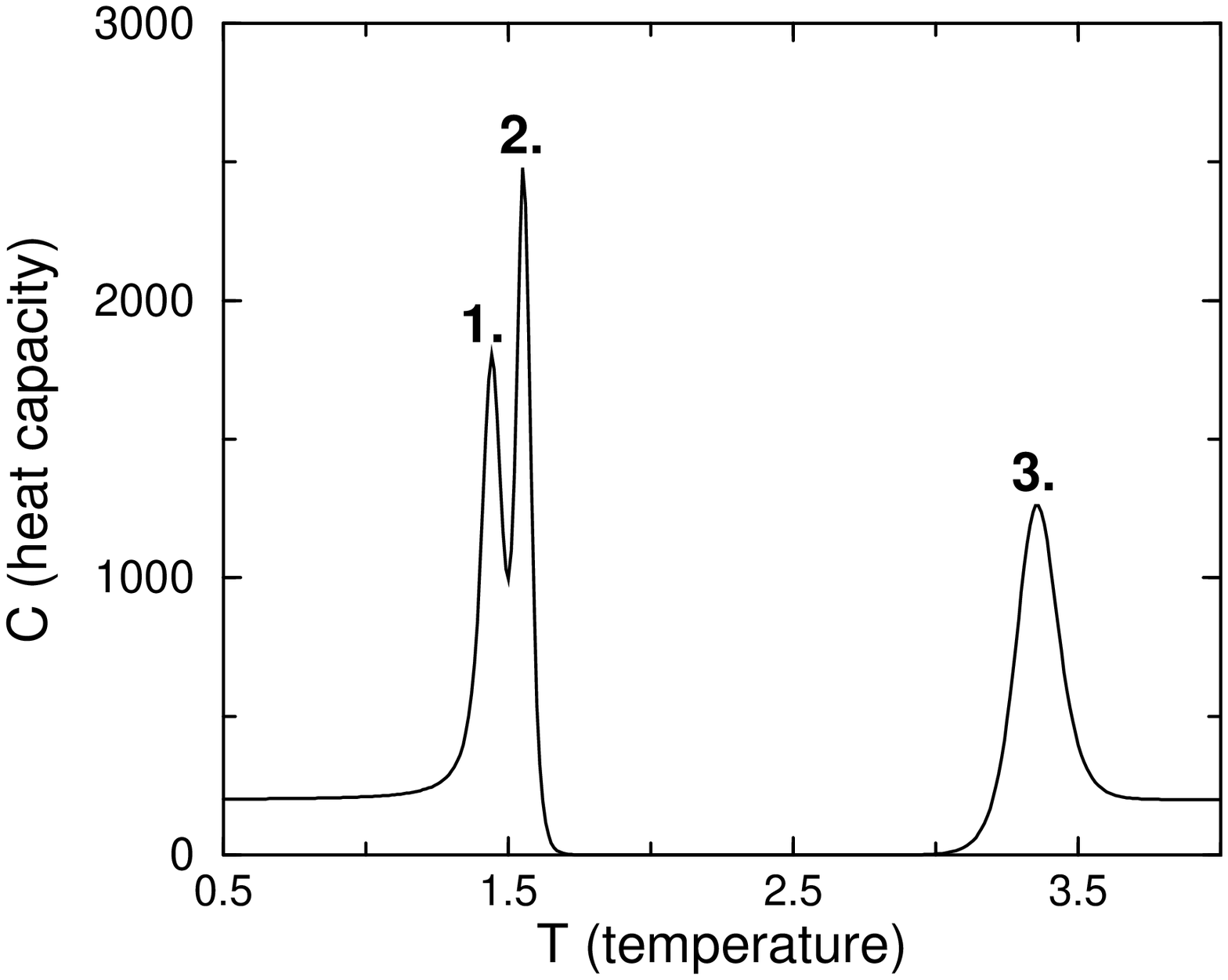,width=\linewidth}}
   \label{fig:4}
\end{figure}

\vspace{0.5cm}
\noindent
{\bf Fig. 3.} The corresponding order parameter $n(T)$ to Fig.\,\ref{fig:2}, describing the degree of folding. The chosen parameters are as in Fig.\,\ref{fig:2}. $n=0$ corresponds to an unfolded protein, while $n=1$ is interpreted as a completely folded protein.  

\vspace{0.5cm}
\noindent
{\bf Fig. 4.} Heat capacity where the effective chemical potential is $\mu = 3.2$, slightly decreased from the value in Fig.\,\ref{fig:2} ($\mu =3.3$). All other parameters are chosen as in Fig.\,\ref{fig:2}. The qualitative picture is a broader separation between $T_2$ and $T_3$ compared to Fig.\,\ref{fig:2}.

\newpage
\newpage

\vspace*{1cm}
\centering{A. Bakk,  A. Hansen and K. Sneppen}\\
\centering{\it A protein model exhibiting three folding transitions}\\

\newpage

\vspace*{1cm}
\centering{A. Bakk,  A. Hansen and K. Sneppen}\\
\centering{\it A protein model exhibiting three folding transitions}\\

\newpage

\vspace*{1cm}
\centering{A. Bakk,  A. Hansen and K. Sneppen}\\
\centering{\it A protein model exhibiting three folding transitions}\\

\newpage

\vspace*{1cm}
\centering{A. Bakk,  A. Hansen and K. Sneppen}\\
\centering{\it A protein model exhibiting three folding transitions}\\

\end{document}